\documentclass[a4paper]{article}

\usepackage{INTERSPEECH2019}
\usepackage{amsmath,graphicx,indentfirst,url,dirtree,color}
\usepackage{ascmac}
\usepackage{CJKutf8}

\newcommand{\Table}[1]{Table \ref{tb:#1}} 
\newcommand{\jp}[1]{ \begin{CJK}{UTF8}{ipxm}#1\end{CJK} }
\newcommand{\br}[1]{\left[#1\right]} 
\newcommand{\revise}[1]{\textcolor{black}{#1}}

\title{JSSS: free Japanese speech corpus for summarization and simplification}

\name{Shinnosuke Takamichi$^1$, Mamoru Komachi$^2$, Naoko Tanji$^1$, and Hiroshi Saruwatari$^1$}
\address{$^1$Graduate School of Information Science and Technology, The University of Tokyo, Tokyo, Japan \\
$^2$Graduate School of \revise{Systems} Design, Tokyo Metropolitan University, Tokyo, Japan}

\email{shinnosuke\_takamichi@ipc.i.u-tokyo.ac.jp}

\begin{document}
\maketitle
\begin{abstract}
    In this paper, we construct a new Japanese speech corpus for speech-based summarization and simplification, ``JSSS'' (pronounced ``j-triple-s''). Given the success of reading-style speech synthesis from short-form sentences, we aim to design more difficult tasks for delivering information to humans. Our corpus contains voices recorded for two tasks that have a role in providing information under constraints: duration-constrained text-to-speech summarization and speaking-style simplification.  It also contains utterances of long-form sentences as an optional task. This paper describes how we designed the corpus, which is available on our project page.
\end{abstract} \vspace{1mm}


\section{Introduction}
    Text-to-speech (TTS) synthesis achieved to synthesize human-quality speech~\cite{oord16wavenet,wang17tacotron,saito18advss} in very limited tasks (e.g., reading-style speech synthesis from short-form sentences of some rich-resourced languages). Both open-source code and open speech corpora help open innovation of speech-based technologies. Since 2017, we have released high-quality and large-scale Japanese speech corpora. The JSUT and JSUT-song corpora~\cite{sonobe17jsut,jsutsong_corpus} are for speaking-/singing-voice synthesis, and the JVS and JVS-MuSiC corpora~\cite{takamichi19jvs,tamaru20jvsmusic} are for multi-speaker/singer modeling. Open projects~\cite{neutrino,watanabe18espnet,hayashi20espnettts,nnsvs} developed by third parties provide synthesis engines and machine learning recipes using our corpora. 
    
    With the success of reading-style speech synthesis from short-form sentences, we aim to design two challenging tasks for delivering information to humans: 1) duration-constrained text-to-speech summarization and 2) speaking-style simplification. The former summarizes text by a spoken language at a desired duration, and the latter synthesizes speech intelligible for \revise{non-native speakers}. These tasks help provide information under the constraints of time limitations or language proficiency. They are challenging because their speech characteristics are far from those of basic reading-style speech.
    
    For these tasks, we developed a new Japanese speech corpus, \textit{JSSS} (pronounced ``j-triple-s''). Our corpus composes speech data and its transcription. We recorded speech with high-quality settings: studio-recording, uncompressed audio format, and a well-experienced native speaker. We also recorded speech of short- and long-form sentences as an optional task. Our corpus has eight hours of high-quality speech data and is available at our project page~\cite{jsss_corpus}. From the next section, we describe how we designed the corpus.
 
\section{Corpus design}
    Our corpus consists of the following four sub-corpora. 
        \begin{itemize} \leftskip -7mm \setlength{\itemsep}{-1pt}
            \item[] \textbf{summarization}: 125 utterances for duration-constrained text-to-speech summarization
            \item[] \textbf{simplification}: 184 short utterances spoken in slow, intelligible style
            \item[] \textbf{short-form}: 3284 short utterances spoken with read style
            \item[] \textbf{long-form}: 168 long utterances spoken with read style
        \end{itemize}
    The directory structures of the corpus are listed below. \textit{[SUB\_DIR\_NAME]} indicates the sub-directory described in the following sections.
                \dirtree{%
        .1 \includegraphics[width=0.25cm]{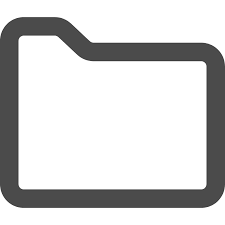} \textbf{summarization}.
            .2 \includegraphics[width=0.25cm]{fig/dir.png} wav24kHz16bit. 
            .2 \includegraphics[width=0.25cm]{fig/dir.png} original\_utf8.  
            .2 \includegraphics[width=0.25cm]{fig/dir.png} transcript\_utf8.
        .1 \includegraphics[width=0.25cm]{fig/dir.png} \textbf{simplification}.
            .2 \includegraphics[width=0.25cm]{fig/dir.png} wav24kHz16bit. 
            .2 \includegraphics[width=0.25cm]{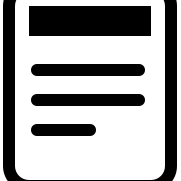} transcript\_utf8.txt.
            .2 \includegraphics[width=0.25cm]{fig/file.png} hiragana\_utf8.txt.
        .1 \includegraphics[width=0.25cm]{fig/dir.png} \textbf{short-form}.
            .2 \includegraphics[width=0.25cm]{fig/dir.png} SUB\_DIR\_NAME.  
                .3 \includegraphics[width=0.25cm]{fig/dir.png} wav24kHz16bit.
                .3 \includegraphics[width=0.25cm]{fig/file.png} transcript\_utf8.txt.
            .2 \includegraphics[width=0.25cm]{fig/dir.png} .... 
        .1 \includegraphics[width=0.25cm]{fig/dir.png} \textbf{long-form}.
            .2 \includegraphics[width=0.25cm]{fig/dir.png} SUB\_DIR\_NAME.
                .3 \includegraphics[width=0.25cm]{fig/dir.png} wav24kHz16bit. 
                .3 \includegraphics[width=0.25cm]{fig/dir.png} original\_utf8.
                .3 \includegraphics[width=0.25cm]{fig/dir.png} transcript\_utf8.
            .2 \includegraphics[width=0.25cm]{fig/dir.png} .... 
        }

    \subsection{Summarization}
        Automatic text summarization generates a short, coherent summary of given text~\cite{dorr03hedgetrimmer,banko00statisticalsumamrization}, shortening it while retaining its important content. Text-length-constrained text summarization~\cite{makino19textsummarizationlength} is text summarization technology that has practical application; it abstractively summarizes text to fit a device that displays a summary~\cite{saito20lengthcontrollablesummarization}. Against such \textit{textual} length constraints, this sub-section addresses \textit{speech} length constraint. Namely, we propose a new task named \textit{speech-length-constrained} or \textit{duration-constrained} text-to-speech summarization. It abstractively summarizes text with a spoken language to fit a desired speech duration.   
        
        We recorded speech for this task. The text to be summarized was web news, which we saved in original\_utf8/*.txt. Our speaker summarized the texts and uttered them to fit duration that the speaker chose in advance. The durations chosen for each text were around 30 and 60 sec. We did not set time limits for recording, and the speaker could re-record as many times as needed. After the recording, first, we manually transcribed the speech. Then, we manually added punctuation at the phrase breaks and added sentence-level time alignment as shown below. We saved the transcription in transcript\_utf8/*.txt.
                            \begin{screen} 
                    \footnotesize
                    $\br{\textrm{START\_TIME}}$ $\br{\textrm{FINISH\_TIME}}$ $\br{\textrm{TRANSCRIPTION}}$ \\
                    00.000    16.006    \jp{株式会社ベネッセコーポレーションが、20歳から40歳の既婚女性に対して、今年を表す漢字は、というアンケートを行った結果、1位には、おかしい、チェンジという意味を持つ、変が選ばれました。} \\ 
                    17.240    26.460    \jp{更に、来年の漢字では、1位が明るい、2位楽しい、3位幸せと、ポジティブな文字が続きました。} \\
                    27.613    32.426    \jp{来年こそは明るく楽しく幸せな1年にとの願いが感じられます。} \\
                    cf.) \\
                    00.000    16.006    Benesse Corporation conducted a survey of married women between the ages of 20 and 40 to find out what the Chinese characters for this year would be, and the number one choice was \jp{変}, which means ``funny'' or ``change.''\\ 
                    17.240    26.460    Furthermore, \revise{positive characters are listed} in the Chinese characters for ``next year.'' First place is \jp{明るい} (bright), second place is \jp{楽しい} (fun), and third place is \jp{幸せ} (happy). \\
                    27.613    32.426    I feel hopeful that next year will be bright, fun, and happy.
                \end{screen}	 
   
    \subsection{Simplification}
        Given the effect the global pandemic has had on Japan in 2020, there is a question of how we can convey emergency and lifeline information to the approximately three million foreign residents living in Japan~\cite{immigrationjapan}. \revise{The Immigration Services Agency of Japan and the Agency for Cultural Affairs reported that many foreign residents prefer \textit{simple Japanese} rather than English for information services~\cite{simplejapaneseguideline}.} ``Simple Japanese'' speech is much different from standard reading-style speech. ``Simple Japanese'' sentences use daily-use phrases with a limited vocabulary and are uttered in a slow, intelligible style~\cite{shibata07nhkreport}. Text simplification with lexical constraint~\cite{nishihara19textsimplification} can potentially artificially simplify vocabulary in text. On the other hand, here we deal with \textit{speaking-style simplification}, which aims to artificially synthesize speech in a slow, intelligible style. Therefore, we instructed a speaker on the speaking style and recorded speech of simple, pre-designed sentences. An example is below.
                            \begin{screen}
                    \jp{おおきい じしんが おきました} \\
                    cf.) There was a big earthquake.
                \end{screen}
        
        In this sub-corpus, we saved the text in transcript\_utf8.txt and manually converted it into hiragana\footnote{Japanese syllabary that shows pronunciation} in hiragana\_utf8.txt.  
     
    \subsection{Short-form}
        Synthesizing an isolated short utterance is a basic TTS task. To build a basic TTS system, we recorded speech data of short-form sentence utterances. We prepared three subsets (corresponding to \textit{[SUB\_DIR\_NAME]}) from the JSUT corpus~\cite{sonobe17jsut} as follows.
            \begin{itemize} \leftskip -5mm \itemsep -1pt
                \item[] \textbf{voiceactress100}: phoneme-balanced minimal set
                \item[] \textbf{onomatopee300}: mid-sized set that includes Japanese onomatopoeias 
                \item[] \textbf{basic5000}: large-sized set
            \end{itemize}
            
        After the recording, we manually added punctuation at the phrase breaks. Note that, the positions are different from the original data stored in the JSUT corpus.
        
    \subsection{Long-form}
        When uttering a long-form sentence that consists of multiple sentences, human speakers usually insert phrase breaks between word transitions without punctuation. This plays an important role in listenable and expressive speech. Synthesizing such speech is more challenging than synthesizing basic short utterances. To construct a corpus for it, we recorded speech uttering Wikipedia articles~\cite{wikipedia} (corresponding to \textit{[SUB\_DIR\_NAME]}). Our speaker uttered articles paragraph by paragraph, excluding tables, figures, and their captions. After the recording, we manually added punctuation at the phrase breaks and sentence-level time alignment as shown below.
                            \begin{screen}
                    $\br{\textrm{START\_TIME}}$   $\br{\textrm{FINISH\_TIME}}$   $\br{\textrm{TRANSCRIPTION}}$ \\
                    0.347	13.144     \jp{香川県において、うどんは地元で特に好まれている料理であり、一人あたりの消費量も、日本全国の都道府県別統計においても、第1位である。} \\ 
                    14.574	32.954    \jp{料理等に地域名を冠してブランド化する地域ブランドの1つとしても、観光客の増加、うどん生産量の増加、知名度注目度の上昇などの効果をもたらし、地域ブランド成功例の筆頭に挙げられる。} \\
                    ... \\
                    cf.) \\
                    0.347	13.144   In Kagawa Prefecture, udon is a particularly popular local dish, and the amount consumed per person is also the highest in Japan in terms of prefectural statistics. 
                    14.574	32.954   This is one of the most successful examples of a regional brand using the name of a region as a brand for food and other items, resulting in an increase in the number of tourists, an increase in the amount of udon produced, and an increase in name recognition. \\
                    ...
                \end{screen}
            
        In this sub-corpus, we saved the original text in original\_utf8/*.txt and transcription in transcript\_utf8/*.txt. Note that, punctuation in the transcribed text was inserted at the phrase breaks, so the positions differ from that of the original text.
	        
\section{Results of data collection}	
        \begin{table*}[t]
\centering
\caption{Statistics of sub-corpora.}
\label{tb:statistics}
\begin{tabular}{|c||c|c|c|c|c|c|}
\hline
Sub-corpus          & Style         & \#utterances      & Duration [hour]   & Duration / utt. [sec]  \\ \hline\hline
summarization       & news-reading  & 125               & 1.69              & 48.8               \\ \hline
simplification      & slow-speaking & 184               & 0.26              & 5.09               \\ \hline
short-form          & reading       & 3284              & 4.03              & 4.42               \\ \hline
long-form           & reading       & 168               & 2.35              & 50.4              \\ \hline\hline
Total               & -             & 3761              & 8.33              & 7.97               \\ \hline
\end{tabular}
\end{table*}

	\subsection{Settings}
	    We hired a female native Japanese speaker who is not a professional speaker but has voice training. We recorded her voice in an anechoic room at the University of Tokyo using an iPad mini with a mounted SHURE MV88A-A microphone. The first author directed the recording. Her voice was originally sampled at 48~kHz and downsampled to 24~kHz by SPTK \cite{sptk}. We recorded in 24-bit/sample RIFF WAV format and encoded in 16-bit/sample format. Sentences (transcriptions) were encoded in UTF-8. For duration-constrained text-to-speech summarization, we used the Livedoor New Corpus~\cite{livedoornewscorpus} as the original text to be summarized. For speaking-style simplification, we followed text and speaking-style instructiond provided by Hirosaki University\footnote{
            We downloaded them from \url{http://human.cc.hirosaki-u.ac.jp/kokugo.html}, but currently they are not available because the laboratory was closed in 2020. 
        }. For long-form utterances, we used three featured Japanese articles from Wikipedia: Sanuki udon (wheat-flour noodles of Japanese cuisine), Masakazu Katsura (Japanese manga artist), and Washington, D.C. (capital of the United States). 

	\subsection{Statistics}
	    \Table{statistics} lists the number of utterances and durations for each sub-corpus. ``Simplification'' and ``short-form'' consist of short utterances approximately 5 seconds per utterance. ``Summarization'' and ``long-form'' consist of utterances approximately 50 seconds per utterance, approximately 10 times longer than short utterances. The total duration is approximately 8 hours, which is slightly shorter than our previous corpus~\cite{sonobe17jsut} designed for the end-to-end TTS.
	    
\section{Conclusion}	
    In this paper, we constructed the JSSS voice corpus. We designed the corpus for text-to-speech summarization, speaking-style simplification, and short-/long-form TTS synthesis. 

\section{License}
    The text files are licensed as below.
        \begin{itemize} \itemsep -1mm
            \item summarization/ ... CC BY-ND 2.1~\cite{livedoornewscorpus}
            \item simplification/ ... No commercial use
            \item short-form/ ... CC BY-SA 4.0 etc~\cite{jsut_corpus}
            \item long-form/ ... CC BY-SA 4.0
        \end{itemize}
    The speech files may be used for 
    \begin{itemize} \itemsep -1mm
        \item Research by academic institutions
        \item Non-commercial research, including research conducted within commercial organizations
        \item Personal use, including blog posts.
    \end{itemize}

\section{Acknowledgements}
    Part of this work was supported by the GAP foundation program of the University of Tokyo. We thank RONDHUIT for allowing us to distribute the text-to-speech summarization sub-corpus. We thank Mr. Takaaki Saeki, Mr. Yota Ueda, and Mr. Taiki Nakamura of the University of Tokyo for their help.

\ninept
\bibliographystyle{IEEEbib}
\bibliography{tts}

\end{document}